\documentclass[aps,prb,reprint,noshowkeys,superscriptaddress]{revtex4-1}
\usepackage{graphicx,bm,xcolor,amscd,amsmath,amssymb,amsfonts,physics,wrapfig,txfonts,siunitx,dcolumn}
\usepackage{subfig,comment}

\usepackage[version=4]{mhchem}
\usepackage[nolist]{acronym}
\usepackage{soul}

\begin{document}
\newcommand{\basis}[0]{\mathcal{B}}
\newcommand{\hvmc}[0]{H_{\text{VMC}}}
\newcommand{\hvmcb}[0]{\hat{H}_{\text{VMC}}^\basis}
\newcommand{\htc}[0]{H_{\text{TC}}}
\newcommand{\htcb}[0]{\hat{H}_{\text{TC}}^\basis}
\newcommand{\obasis}[0]{\mathcal{B}^{\perp}}
\newcommand{\vbasis}[0]{\mathcal{V}_\mathcal{B}}
\newcommand{\vobasis}[0]{\mathcal{V}_{\mathcal{B}}^\perp}
\newcommand{\pbasis}[0]{\hat{P}_{\mathcal{B}}}
\newcommand{\pobasis}[0]{\hat{P}_{\mathcal{B}}^\perp}
\newcommand{\ident}[0]{\mathbf{1}}

\newcommand{\mt}[1]{\multicolumn{1}{c}{#1}}
\newcommand{\manu}[1]{{\color{red}#1}}
\newcommand{\br}{\mathbf{r}}
\newcommand{\brs}[1]{\mathbf{r}_{#1}}
\newcommand{\Ndet}{N_{\text{det}}}
\newcommand{\Htc}{\hat{H}_{\text{TC}}}
\newcommand{\Hnu}{\hat{H}_{\nu}}
\newcommand{\Hv}{\hat{H}_{V}}
\newcommand{\Var}{\mathrm{Var}}
\newcommand{\Cov}{\mathrm{Cov}}
\newcommand{\kmu}[2]{\hat{K}[\mu](\brs{#1},\brs{#2})}
\newcommand{\kone}[2]{\hat{V}[\{ \beta_A\}](\brs{#1},\brs{#2})}
\newcommand{\deriv}[3]{\frac{\partial^{#3} #1}{\partial {#2}^{#3}}}
\newcommand{\lmu}[3]{\hat{L}[\mu](\brs{#1},\brs{#2},\brs{#3})}
\newcommand{\vone}[1]{\hat{v}[\{ \beta_A\}](\brs{#1})}


\newcommand{\LCPQ}{Laboratoire de Chimie et Physique Quantiques (UMR
  5626), Université de Toulouse, CNRS, UPS, France}

\newcommand{\LCT}{Laboratoire de Chimie Th\'eorique (UMR
  7616), Université Paris Sorbonne, CNRS, France}

\title{Optimization of large determinant expansions in quantum Monte Carlo}

\author{Abdallah \surname{Ammar}}
\affiliation{\LCPQ}

\author{Emmanuel \surname{Giner}}
\affiliation{\LCT}

\author{Anthony \surname{Scemama}}
\email{scemama@irsamc.ups-tlse.fr}
\affiliation{\LCPQ}

\begin{abstract}
We present a new method for the optimization of large configuration
interaction (CI) expansions in the quantum Monte Carlo (QMC)
framework.  The central idea here is to replace the non-orthogonal
variational optimization of CI coefficients performed in usual QMC
calculations by an orthogonal non-Hermitian optimization thanks to the
so-called transcorrelated (TC) framework, the two methods yielding the
same results in the limit of a complete basis set.
By rewriting the TC equations as an effective self-consistent Hermitian
problem, our approach requires the sampling of a single quantity per Slater
determinant, leading to minimal memory requirements in the QMC code.
Using analytical quantities obtained from both the TC framework and
the usual CI-type calculations, we also propose improved estimators which
reduce the statistical fluctuations of the sampled quantities by more
than an order of magnitude.
We demonstrate the efficiency of this method on wave functions containing
$10^5-10^6$ Slater determinants, using effective core potentials or
all-electron calculations. In all the cases, a sub-milliHartree convergence is
reached within only two or three iterations of optimization.
\end{abstract}

\maketitle


\begin{acronym}
  \acro{BLM}{blocked linear method}
  \acro{CAS}{complete active space}
  \acro{CC}{coupled cluster}
  \acro{CI}{configuration interaction}
  \acro{CISD}{\ac{CI} with single and double substitutions}
  \acro{DFT}{density functional theory}
  \acro{ECP}{effective core potential}
  \acro{FCI}{full configuration interaction}
  \acro{FCIQMC}{full configuration interaction quantum Monte Carlo}
  \acro{GD}{gradient descent}
  \acro{GEP}{generalized eigenvalue problem}
  \acro{HF}{Hartree-Fock}
  \acro{LM}{linear method}
  \acro{MO}{molecular orbital}
  \acro{QMC}{Quantum Monte Carlo}
  \acro{SCF}{self consistent field}
  \acro{SR}{stochastic reconfiguration}
  \acro{TC}{transcorrelated}
  \acro{VMC}{variational Monte Carlo}
  \acro{ZV}{zero-variance}
\end{acronym}

\section{Introduction}

The accurate description of highly correlated systems such as
transition states, magnetic systems and excited states requires a
multi-configurational form of the wave function.
For most problems, the \ac{CAS} picture already contains enough
information to give a qualitatively correct description of the physics
governing the electronic structure.
There exists nevertheless some systems in which the dynamical
correlation is strongly coupled to the static correlation,\cite{CalAngTarCabMal-JCP-09,Gozem-retinal-JCTC-12,MalrCabCalGraGui-CR-14,liu_2014,hapka_2020}
such that the dominant part of the wave function (typically its projection onto a valence-like \ac{CAS})
is severely impacted by the full treatment of dynamical correlation in the optimization process.
Multiple methods were designed to take into account the feedback of the
dynamical correlation on the reference: one can cite for instance internally decontracted
\ac{CI}~\cite{angeli_2006,limanni_2013} or
\ac{CC}~\cite{meller_1996,mahapatra_1999}
methods, $f_{12}$ methods combined
with multi-reference \ac{CI},~\cite{shiozaki_2011} range-separated \ac{DFT} combined with
\ac{CAS},~\cite{savin_1996, sharkas_2012, pernal_2022} or the
shifted-$B_k$ method~\cite{nitzsche_1978, garniron_2018}.

In this paper, we focus on \ac{QMC} approaches, in which the
$N$-electron wave function $\Psi$ is expressed as
\begin{equation}
 \label{eq:def_psi}
    \Psi(\br) = \Phi(\br) \, e^{J(\br)},
\end{equation}
where $\br$ is the $\mathbb{R}^{3N}$ electronic configuration space.
In Eq.~\eqref{eq:def_psi}, $\Phi$ is a multi-determinant expansion and $\exp(J)$ is a
Jastrow correlation factor, taking explicitly into account
electron-electron distances.
The Jastrow factor allows the wave function $\Psi(\br)$ to fulfill the exact
Kato's cusp conditions~\cite{Kato_1957}, but also introduces a large amount of short-range dynamical correlation
which are usually very costly to capture within a usual Slater determinant expansion.
Therefore, the functional form of Eq.~\eqref{eq:def_psi} is very flexible and can be used to treat
complex correlation effects, provided that one manages to optimize the different parameters.

As opposed to $f_{12}$ methods where the correlation factor is
projected into a space orthogonal to the multi-determinant expansion,
the Jastrow factor has a significant overlap with the determinant
expansion. This last statement particularly motivates the need for an
optimization of the parameters of $\Phi$ in the presence of the
Jastrow factor, which is not straightforward as it involves
$3N-$dimensional integrals that cannot be evaluated exactly in the
general case. As a consequence, the necessary matrix elements are
sampled in a \ac{VMC} simulation and they are subject to statistical noise.
In the general context of \ac{VMC} simulations, an important aspect
is precisely the amplitude of the statistical fluctuations of the quantities
needed to optimize the wave function, which eventually determines the actual
applicability of a given computational algorithm within a reasonable CPU time.

One of the most common methods to perform such an optimization is
the standard \ac{LM}, where the Schrödinger equation is projected in the self-plus-tangent space,
i.e., the space spanned by the current wave function and its
first derivatives with respect to the $N_p$
parameters.~\cite{nightingale_2001,umrigar_2007,toulouse_2007,toulouse_2008}
The new parameters are obtained by solving a \ac{GEP}
after building the  Hamiltonian and overlap matrices,
both of size $N_p \times N_p$, through a \ac{VMC} calculation.
In practice, the standard \ac{LM} is limited to few thousands of
parameters~\cite{clark_2011}
due to the large memory requirement to sample and store these matrices.

Another class of optimization methods constrained by the storage of large matrices
is the \ac{SR} method.~\cite{sorella_2001,casula_2004,sorella_2007}
In this method, the imaginary time evolution operator is expanded to first order
and projected iteratively in the self-plus-tangent space.
At each step of this procedure, one must build a $N_p \times N_p$
overlap-like matrix and solve a system of linear equations.
The wave function is updated after each iteration until the energy has converged.

Several solutions have been proposed to address the memory bottleneck of
the \ac{LM} and \ac{SR} methods.
Neuscamman and co-workers~\cite{neuscamman_2012,neuscamman_2016}
have suggested to employ Krylov subspace solvers to avoid building the matrices explicitly.
Although this approach allows to enlarge significantly the number of variational parameters in
the optimization, it requires additional matrix-vector multiplications as
the Hamiltonian and overlap matrices are sampled, which increases
the sampling effort.
Moreover, ill-conditioned matrices constitute a challenge in the
\ac{LM}.~\cite{zhao_2017}
A similar strategy relies on the Jacobi-Davidson method,~\cite{sleijpen_1996,sleijpen_2000,sabzevari_2020}
which generalizes Davidson's method.~\cite{davidson_1975}
Recently, a variant of the \ac{LM}, termed the \ac{BLM}, has been introduced
to alleviate the difficulties of storing large matrices.~\cite{zhao_2017}

At the other end of the scale, the \ac{GD} approaches~\cite{harju_1997}
have a smaller memory footprint as  the required storage
scales linearly with the number of parameters.
These approaches exploit the recent developments
in the field of deep-learning algorithms of neural networks
and make use of some adaptive \ac{GD} flavours
to perform the optimization directly.~\cite{schwarz_2017,sabzevari_2018,luo_2019,mahajan_2019,mcfarland_2022}
In their recent work\cite{otis_2019}, Otis and Neuscamman have presented a comparison between
\ac{LM} and \ac{GD} techniques, and they have developed a hybrid scheme
which combines \ac{BLM} and \ac{GD} methods.

In this work we focus on the optimization of the \ac{CI} coefficients,
in view of optimizing very large \ac{CI} expansions for an arbitrary choice of Jastrow factor.
For this purpose, we present an iterative scheme to optimize the wave function by
combining \ac{VMC} and the \ac{TC} approach, motivated by i) the fact
that the \ac{TC} formalism doesn't require the sampling of the overlap matrix
and ii) that the \ac{VMC} and \ac{TC} approaches lead to the same wave
function in the limit of a complete basis set.

In the \ac{TC} framework the correlation effects brought by the Jastrow factor are incorporated directly in the
original Hamiltonian through a similarity transformation $\Htc \equiv e^{-\hat{J}} \hat{H} e^{\hat{J}}$.
This approach, originally proposed by Hirschfelder to
remove electron-electron poles from the original Hamiltonian,\cite{hirschfelder_1963}
was then further developed by Boys and Handy who referred to it as a
\emph{\ac{TC} method} where the authors optimize both the Jastrow factor and the orbitals of a single Slater determinant.~\cite{boys_1969_I,boys_1969_II,boys_1969_III,boys_1969_IV,handy_1971,handy_1972}
The \ac{TC} method was then further developed in the beginning of the 2000's by Ten-No\cite{ten_2000_I,ten_2000_II,Hino_2001} and co-workers through
the use of a universal correlation factor (\textit{i.e.} the same correlation factor for all systems)
with a relatively flexible form for the wave function expansion (such as perturbation theory or linearized coupled cluster).
Developments using \ac{VMC} to compute the variance of the \ac{TC} Hamiltonian in order to remain within a variational framework
were carried by Umezawa and co-workers
\cite{umezawa_2003,Umezawa_2003_I,Umezawa_2003_II,Umezawa_2004,Umezawa_2005,Prasad_2007}.
A recent renewal of the \ac{TC} methods were brought by Alavi and coworkers where they used a flexible form
for both the correlation factor and the wave function\cite{Luo_2018,Liao_2021,Schraivogel_2021} thanks
to the use of a version of the \ac{FCIQMC}\cite{Booth_2009,Cleland_2010} method adapted for a non-Hermitian Hamiltonian containing up to three-electron interactions.

The algorithm exposed here proposes to bypass the usual \ac{VMC}
optimization, and use instead the \ac{TC} method to optimize the
\ac{CI} coefficients of Jastrow-Slater wave functions for an arbitrary
Jastrow factor.  To do so, we reformulate the non-Hermitian \ac{TC}
approach in terms of an Hermitian self-consistent dressing of the
standard Hamiltonian $\hat{H}$ which accounts for the effect of the
Jastrow factor.  These equations are then projected in a basis of
Slater determinants, and as the scheme is Hermitian, the standard
Davidson algorithm can be employed to optimize the wave function.  The
advantage of the present scheme is that one samples the action of the
dressed Hamiltonian on the \ac{CI} vector, which requires the sampling
of a single quantity per \ac{CI} coefficient, leading to a minimal
memory footprint.  The convergence to the chemical accuracy of the
self-consistent procedure is reached with typically two or three
iterations.  Another interesting aspect of the present scheme is that
the sampled dressing matrix has a zero-variance property which leads
to relatively small fluctuations as compared to the estimators used in
the \ac{LM}, and the fluctuations can be further reduced by
introducing the deterministic computation of an auxiliary \ac{TC}
Hamiltonian.

The paper is organized as follows.
In section~\ref{sec:opt_theory}, we discuss the most common
schemes used to optimize \ac{CI} coefficients in the presence of a Jastrow factor.
We then give in Sec.~\ref{sec:tc_for_vmc} the description of our new algorithm:
in Sec.~\ref{subsec:tc_for_vmc} we provide a brief description of the \ac{TC} formalism and its connection with \ac{VMC},
we present the general ideas of our optimization scheme for large \ac{CI} expansions in Sec.~\ref{sec:iterative}
and present in Sec.~\ref{subsec:dress_elements} the different numerical strategies employed to compute
the dressing elements with minimal statistical fluctuations.
In Sec.~\ref{sec:opt_numeric}, we present numerical tests validating the present approach.
In Sec.~\ref{sec:Be} we test the feasibility of the present approach on the Beryllium atom
together with the actual impact of the incompleteness of the $N$-electron basis set.
Finally, in Sec.~\ref{sec:mol} we optimize several \ac{CI} expansions for small molecules made of
a few hundred thousand Slater determinants.

\section{Wave function optimization in the presence of a Jastrow factor}\label{sec:opt_theory}
\subsection{General context}
Consider a ground-state $N$-electron wave function $\Phi$ expressed in a basis of
Slater determinants $\basis=\{D_I, 1 \le I \le \Ndet\}$ obtained with orthonormal spin-orbitals
\begin{equation}
 \label{eq:phi_di}
  \Phi(\br) = \sum_{I=1}^{\Ndet} c_I \, D_I(\br),
\end{equation}
where $\Ndet$ is the number of determinants.
In the \ac{QMC} framework, a relatively cheap and efficient way of
increasing the amount of correlation energy described by the wave
function is to introduce a Jastrow factor $J(\br)$, which captures short-range effects
that cannot be easily described by the finite determinant basis set
\begin{equation}
  \Psi(\br) = \Phi(\br) \, e^{J(\br)} = \sum_{I=1}^{\Ndet} c_I\, D_I(\br)\, e^{J(\br)},
  \label{eq:ansatz}
\end{equation}
where, generally,
$J(\br)$ is a function of electron-nuclear, electron-electron, and electron-electron-nuclear distances.
Because of the large overlap between the Jastrow factor and the determinant basis $\basis$,
the optimal \ac{CI} coefficients are not those obtained by simply minimizing the variational energy
of the wave function of Eq.~\eqref{eq:phi_di}.
Our goal in this work is to implement an efficient scheme to optimize the \ac{CI} coefficients for large $\Ndet$ in
the presence of a general Jastrow factor $J(\br)$.
Two methods widely used to perform such an optimization, the \ac{LM} and \ac{SR}, are briefly discussed in this section.

\subsection{Linear method}

The most natural way to optimize the \ac{CI} coefficients is to
express the \ac{CI} problem in the basis of determinants augmented by
a Jastrow factor $\mathcal{B}_J=\qty{D_I\,e^{J}, 1 \le I \le \Ndet}$.
This basis is not orthonormal, so in addition to the Hamiltonian
matrix elements
\begin{equation}
  H_{IK} = \mel{D_I\,e^{J}}{\hat{H}}{D_K\,e^{J} },
\end{equation}
one needs also to compute the overlap matrix elements
\begin{equation}
  S_{IK} = \braket{D_I\,e^{J}}{D_K\,e^{J}}.
\end{equation}
As the forms commonly used for the Jastrow factor are too complicated
to integrate analytically, these $3N$-dimensional integrals need to be sampled using a
\ac{VMC} sampling:
\begin{align}
  \label{eq:hij}
  H_{IK} &= \left \langle \frac{D_I \, e^{J}}{\Psi} \frac{\hat{H}\, \left(D_K \, e^{J} \right)}{\Psi}
  \right \rangle_{\Psi^2}, \\
  \label{eq:sij}
  S_{IK} &=  \left \langle \frac{D_I \, e^{J}}{\Psi} \frac{D_K \, e^{J}}{\Psi}
  \right \rangle_{\Psi^2},
\end{align}
where $\langle \dots \rangle_{\Psi^2}$ denotes the stochastic average
over the Monte Carlo samples drawn with the $3N$-dimensional density
$\Psi^2$.

The \ac{CI} problem can now be solved as a \ac{GEP}
\begin{equation}
 \label{eq:general_h}
  \mathbf{H \, C} = E \mathbf{S\,C}.
\end{equation}
Provided that the Jastrow factor is kept constant, solving this \ac{GEP} provides directly the best variational
coefficients in the basis $\mathcal{B}_J$. However, several difficulties are encountered in this approach.
The statistical errors obtained from the \ac{VMC} calculation on
$H_{IK}$ and $S_{IK}$ can be important, and
frequently larger than the matrix element itself, leading
to an effective optimization of only a reduced set of the parameters of the wave function.
Therefore, this approach is in practice limited by the expensive memory requirements
for the storage of the sampled matrices $\mathbf{H}$ and $\mathbf{S}$.

The memory cost scales as $\mathcal{O} \left( \Ndet^2 \right)$,
so when the number of parameters becomes as large as a few thousands
this simple approach becomes prohibitive.
A first solution to this memory bottleneck is to employ a Krylov
subspace solver to eschew building $\mathbf{H}$ and $\mathbf{S}$
explicitly.~\cite{neuscamman_2012} This improvement has lead to
the optimization of up to $5 \times 10^5$ variational parameters.
The \ac{BLM} is an alternative approach~\cite{zhao_2017} in which
the space of determinants $\mathcal{B}_J$ is divided into $N_b$ blocks.
A \ac{GEP} is solved in each block to generate a set
of $N_k$ eigenvectors.
Those eigenvectors are then fixed and employed as directions to find
a new direction for the full space $\mathcal{B}_J$.
The memory cost of this approach scales approximately as
$\mathcal{O}\left(\Ndet^2 / N_b \right)$.
The \ac{BLM} was recently applied in combination with \ac{GD} methods.~\cite{otis_2019}

\subsection{Stochastic reconfiguration}
The \ac{SR} is an alternative method where the wave function is iteratively improved
by applying the first order expansion of the
imaginary time evolution operator
$\exp\left( - \tau \hat{H} \right) \approx \left( 1 - \tau \hat{H}  \right)$. \cite{sorella_2001,casula_2004,sorella_2007}
Instead of solving a \ac{GEP} one has to solve a set of $\Ndet$ linear
equations
\begin{equation}
    \mathbf{\overline{S}} \, \delta \mathbf{c} = -\frac{\tau}{2} \, \mathbf{g},
    \label{eq:SR_lineq}
\end{equation}
where
\begin{align}
  \label{eq:smodifih}
  \overline{S}_{IK} &= S_{IK} -
                       \left \langle \frac{D_I \, e^{J}}{\Psi} \right \rangle_{\Psi^2}
                       \left \langle \frac{D_K \, e^{J}}{\Psi} \right \rangle_{\Psi^2} \\
  \label{eq:gi}
  g_{I} &= 2 \left[
              \left \langle \frac{D_I \, e^{J}}{\Psi} \frac{\hat{H}\,\Psi}{\Psi} \right \rangle_{\Psi^2} -
              \left \langle \frac{\hat{H}\,\Psi}{\Psi} \right \rangle_{\Psi^2}
              \left \langle \frac{D_I \, e^{J}}{\Psi} \right \rangle_{\Psi^2}
              \right].
\end{align}

In its original formulation, the \ac{SR} required storing an overlap-like matrix
of $\Ndet \times \Ndet$ dimension which restricts the optimization to
a few thousand coefficients.
However, this memory bottleneck can be bypassed by using
a conjugate gradient iterative solver to solve Eq.~\eqref{eq:SR_lineq}.~\cite{neuscamman_2012}
In this scheme, the explicit matrix $\mathbf{\overline{S}}$ is not
required, but one needs to store instead Monte Carlo samples over $M$
steps making the storage become $\order{M\times N_p}$. This
improvement allowed to optimize
\ac{CI} expansions of few hundred thousand parameters.~\cite{assaraf_2017,dash_2018}

\section{Using the transcorrelated formalism to approximate VMC}
\label{sec:tc_for_vmc}
In this section, we present a new iterative scheme for the
optimization of large \ac{CI} expansions in the presence of an arbitrary Jastrow factor.
This approach lies in the framework of Krylov subspace solvers.
The memory requirement of the algorithm is minimal and scales as $\mathcal{O}(\Ndet)$,
the convergence within chemical accuracy is reached typically in two or three iterations, and
the method takes advantage of improved estimators which drastically reduce statistical fluctuations.
In Sec.~\ref{subsec:tc_for_vmc} we present the main idea of our approach,
in Sec.~\ref{sec:iterative} we expose the general iterative scheme used to optimize the wave functions,
and in Sec.~\ref{subsec:dress_elements} we detail the different
strategies to reduce the statistical
fluctuations.

\subsection{General idea}
\label{subsec:tc_for_vmc}
In the \ac{TC} formalism, the effect of the Jastrow factor is incorporated in the
Hamiltonian through a similarity
transformation\cite{hirschfelder_1963,boys_1969_I,boys_1969_II,boys_1969_III,boys_1969_IV,handy_1971,handy_1972}
\begin{equation}
    \hat{H}_{J} \equiv  e^{- \hat{J} } \, \hat{H} \,  e^{ \hat{J} }.
\end{equation}
Therefore, solving exactly the Schrödinger equation
\begin{equation}
    \hat{H} \, \Psi = E \, \Psi,
    \label{eq:standard_DE}
\end{equation}
with a wave function $\Psi(\br) = \Phi(\br) \, e^{ J (\br)}$ as
defined in Eq.~\eqref{eq:ansatz}
is equivalent to solve the \ac{TC} eigenvalue equation
\begin{equation}
    \hat{H}_{J} \, \Phi = E \, \Phi,
    \label{eq:TC_DE}
\end{equation}
in the limiting case where $\Phi$ is expanded in a complete $N$-electron basis set.
The effective \ac{TC} Hamiltonian can be written as
\begin{equation}
    \hat{H}_{J}  = \hat{H} + \hat{\Delta}_J,
\end{equation}
with
\begin{equation}
     \hat{\Delta}_J = -\frac{1}{2} \sum_{i=1}^{N}
            \left[ \laplacian_i J + \grad_i J \cdot \grad_i J + 2 \grad_i J \cdot \grad_i \right].
    \label{eq:delta_J}
\end{equation}
Note that the $\hat{\Delta}_J$ operator is non-Hermitian due to the
gradient operator in the last term of Eq.~\eqref{eq:delta_J}. As a consequence each eigenvalue
is associated with a couple of left and right eigenvectors, and the
variational principle does not apply anymore in the sense that there
is no functional $F[\Phi]$ of a $N$-electron $\ket{\Phi}$ satisfying
$F[\Phi] \ge E_0$ where $E_0$ is the exact ground state energy.

The central idea of this paper is to use the \ac{TC} framework to
optimize a \ac{CI} wave function rather than the usual \ac{VMC}
framework. The main technical differences between the \ac{TC} and the
\ac{VMC} approaches are first that in the \ac{TC} framework the
\ac{CI} problem is expressed in an orthonormal basis, and secondly
that the \ac{TC} Hamiltonian involves \textit{at most} three-electron
interactions. Indeed, in the usual \ac{VMC} approach the \ac{CI}
problem is solved in the non-orthonormal basis of determinants
multiplied by the Jastrow factor, and the matrix elements of the
Hamiltonian involve up to $N$-electron interactions.

These two approaches are nevertheless not equivalent when working with an
incomplete basis set. Provided a given Jastrow factor $J$, if one
uses an incomplete set of Slater determinants $\basis$ and the associated projector $\pbasis$ to project
the \ac{TC} equation and the \ac{VMC} equation, one obtains the following projected operators
\begin{equation}
 \htcb = \pbasis\, e^{-\hat{J}} \hat{H} e^{\hat{J}}\,  \pbasis,
\end{equation}
\begin{equation}
 \hvmcb = \pbasis \, e^{+\hat{J}} \hat{H} e^{\hat{J}}\, \pbasis,
\end{equation}
which differ only by the sign in the first exponential.
In the limit of a complete $N$-electron basis, the \ac{TC} equation and the Hermitian non-orthogonal
\ac{GEP} of Eq.~\eqref{eq:general_h} lead to the same eigenvectors and eigenvalues.
However, as shown in the appendix (see Sec.~\ref{sec:appendix}) the
right-eigenvectors of $\htcb$ do not coincide with the eigenvectors
of $\hvmcb$, although being usually very close.
This means that the former is not optimal in the sense of the variational energy.
Nonetheless, the
statistical fluctuations induced by the finite sampling of the matrix
elements of $\hvmcb$ are such that in practice it is very hard to obtain either the exact
optimum in \ac{VMC}, and as we shall see in the Sec.~\ref{sec:Be}, the numerical solutions of
the \ac{TC} and \ac{VMC} problems are close enough to be indistinguishable within reasonable statistical fluctuations.

\subsection{Obtaining right-eigenvectors with a Hermitian dressing of the Hamiltonian}
\label{sec:iterative}
Working in a \ac{TC} framework implies that one needs to rely on
non-Hermitian variants of the Davidson algorithm~\cite{hirao_1982} to obtain the ground
state eigenvector.
In the present section we describe an alternative procedure
to obtain a given eigenstate of the \ac{TC} Hamiltonian involving an effective
non-linear Hermitian Hamiltonian, which can then be easily used with a
standard Hermitian eigensolver.
This idea was initially proposed in the context of single-reference
coupled cluster,~\cite{NebSancMalMay-JCP-95} and further extended to
multi-reference coupled cluster~\cite{GinDavSceMal-JCP-16} and to the
application of the shifted-$B_k$ method to selected
\ac{CI}.~\cite{garniron_2018}

We denote here the projection of the \ac{TC} Hamiltonian in the
basis of Slater determinants $\basis $ by the matrix $\mathbf{\bar{H}}$,
and we use an iterative scheme to obtain a specific right-eigenstate of $\mathbf{\bar{H}}$.
The main idea is to iteratively build a state-specific Hermitian effective Hamiltonian
$\mathbf{\tilde{H}}^{(\Phi)}$ which has, at convergence, the same
eigenvalue and eigenvector as $\mathbf{\bar{H}}$ for the desired state.
Remark that as the present scheme is state-specific, all the other eigenvectors and eigenvalues of
$\mathbf{\tilde{H}}^{(\Phi)}$ are not considered, and hence they are
allowed to be different from those of $\mathbf{\bar{H}}$.

We search for a symmetric dressing matrix $\mathbf{\Delta}^{(\Phi)}$
parameterized by $\Phi$ such that
\begin{equation}
  \mathbf{\tilde{H}}^{(\Phi)}\, \Phi = \qty(\mathbf{H} + \mathbf{\Delta}^{(\Phi)}) \Phi = \mathbf{\bar{H}}\, \Phi.
\end{equation}
The simplest solution is a diagonal dressing matrix:
\begin{equation}
\label{eq:diagonal_dress}
    \mathbf{\Delta}_{II}^{(\Phi)} = \frac{V_I}{c_I} \; ; \;
    \mathbf{\Delta}_{I,J \ne I}^{(\Phi)} = 0
\end{equation}
obtained from the \emph{dressing vector} $V^{(\Phi)}$ defined as
\begin{equation}
    V_{I}^{(\Phi)} = \sum_{K=1}^{\Ndet}  \qty(\bar{\mathbf{H}}_{IK} - \mathbf{H}_{IK} ) \, c_K .
\end{equation}
Iteratively dressing the Hamiltonian using the previously obtained
eigenvector converges to the desired solution.

Nevertheless, using Eq.~\eqref{eq:diagonal_dress} is numerically unstable since the
coefficients $c_I$ can be zero, or extremely small. Instead, we
propose to use a column dressing in column $L$, chosen because $|c_L|$ has the largest
magnitude among all coefficients:
\begin{equation}
    \Gamma_{IL}^{(\Phi)} = \frac{V^{(\Phi)}_I}{c_L} \; ; \;
    \mathbf{\Delta}_{I,J \ne L}^{(\Phi)} = 0
\end{equation}
Then, we build the symmetric matrix $\mathbf{\Delta}^{(\Phi)}$ such that
$\Delta_{LI}^{(\Phi)}=\Delta_{IL}^{(\Phi)} = \Gamma_{IL}^{(\Phi)}$.
Doing this, the effect of the Jastrow factor is counted twice, and
the double-counting can be removed by introducing an extra term in the
diagonal element cancelling out the double counting:
\begin{equation}
  \begin{cases}
    \mathbf{\Delta}_{LL}^{(\Phi)} = 2\, \Gamma_{LL}^{(\Phi)} -
    \sum_{K=1}^{\Ndet}  \Gamma_{KL} c_K/c_L & \\
    \mathbf{\Delta}_{LI}^{(\Phi)} = \mathbf{\Delta}_{IL}^{(\Phi)} = \Gamma_{IL}^{(\Phi)} & I \ne L\\
    \mathbf{\Delta}_{IK}^{(\Phi)} = \mathbf{\Delta}_{KI}^{(\Phi)} = 0 & I\ne L, K \ne L \\
  \end{cases}
\end{equation}
As a result, the dressing matrix $\mathbf{\Delta}^{(\Phi)}$ and thus,
the matrix $\mathbf{\tilde{H}}^{(\Phi)}$ to be diagonalized is symmetric.

To summarize, given a Jastrow factor and a \ac{CI} wave function $\Phi$,
\begin{enumerate}
\item within a \ac{VMC} calculation, we sample the dressing vector $V^{\Phi}$
and the variational energy,
\item we build the matrix elements of $\mathbf{\tilde{H}}^{(\Phi)}$ by
  combining the matrix elements of $\mathbf{H}$ with the sampled
  quantities
\item we employ Davidson's algorithm to extract the ground state right eigenvector of
the matrix $\mathbf{\tilde{H}}^{(\Phi)}$, and we update the wave function $\Phi$.
\end{enumerate}
These steps are iterated until convergence.
Note that only the $\Ndet$ elements $\Gamma_{IL}$ need to be
sampled in \ac{QMC}.
All the other required quantities ($\mathbf{H}_{IK}$) can be obtained with standard
deterministic computational chemistry codes.

\subsection{Calculation of the dressing elements}
\label{subsec:dress_elements}
\subsubsection{Naive estimators}
\label{subsec:naive}
The integrals required to build the dressing elements $\mathbf{\Delta}^{(\Phi)}$ are
\begin{equation}
 \label{eq:dress_0}
  \gamma_{I}^{(\Psi)} = \mel**{D_I}{\hat{H}_{J}}{\Phi} - \mel**{D_I}{\hat{H}}{\Phi},
\end{equation}
and the second term of the right-hand side of Eq.~\eqref{eq:dress_0} can be evaluated analytically for Gaussian-type one-electron
orbitals.
The calculation of the first term of the right-hand side of Eq.~\eqref{eq:dress_0} is however not trivial since it cannot
be calculated in closed form for a general functional form of the Jastrow factor.
It involves three-electron integrals which can be rather expensive and it requires the storage of six-index quantities.
We will show in this section how these integrals can be effectively evaluated in a
\ac{VMC} framework.

The most basic estimator to evaluate the integrals
\begin{equation}
    \alpha_{I}^{(\Psi)} = \mel**{D_I}{\hat{H}_{J}}{\Phi}
    \label{eq:integ_HJ}
\end{equation}
in a \ac{VMC} sampling is
\begin{equation}
    a_I^{(\Psi)}(\br) = \frac{D_I(\br)}{\Phi(\br)} E_{\text{loc}}^{(J)}(\br),
    \label{eq:estimatorI_def}
\end{equation}
where the local energy is defined as
\begin{equation}
     E_{\text{loc}}^{(J)}(\br) = \frac{ \hat{H}_J \, \Phi(\br) }{\Phi(\br)}
                          = \frac{ \hat{H}  \, \left( \Phi(\br)\, e^{J(\br)} \right) }{\Phi(\br) \, e^{J(\br)}},
\end{equation}
and $\Phi^2$ is the density used to draw the
samples of $M$ configurations:
\begin{equation}
    \left \langle a_I^{(\Psi)} \right \rangle_{\Phi^2} \equiv
    \frac{1}{M} \sum_{i=1}^{M} \, a_I^{(\Psi)} \left( \brs{i} \right) \,
    \underset{M\rightarrow \infty}{\longrightarrow }   \alpha_{I}^{(\Psi)}.
\end{equation}

In order to reduce the fluctuations on the dressing elements, we will
present two improved estimators based on the
so-called \emph{control variates}
technique~\cite{MCM_Kalos_Whitlock,caffarel_2019}.
The general idea is to combine the estimator of
Eq~\eqref{eq:estimatorI_def} with a correlated function in the
$3N$-dimensional space of which the integral can be evaluated in a deterministic
way.

\subsubsection{Improved estimator}
\label{subsec:improved}

Consider the modified electronic Hamiltonian
\begin{equation}
    \Hv = \hat{T} + \hat{V}_{\text{n-e}} + \sum_{i<j} V(r_{ij})
    \label{eq:hv_def}
\end{equation}
where $r_{ij} = |\brs{i} - \brs{j}|$, $\hat{T}$ is the usual kinetic operator,
$\hat{V}_{\text{n-e}}$ is the Coulomb interaction between electrons
and nuclei, and
$V(r_{ij})$ is a model potential for electron-electron interaction which allows
an efficient and deterministic evaluation of integrals in a Slater determinant basis
$\mel{D_I}{V}{D_J}$.
The local energy associated with this potential reads
\begin{equation}
    E_{\text{loc}}^{(V)}(\br) = \frac{ \Hv \, \Phi(\br) }{\Phi(\br)}.
\end{equation}
The first improved estimator we propose is
\begin{equation}
    \overline{{a}}_I^{(V)}(\br) = \frac{D_I(\br)}{\Phi(\br)} \,
                        \left( E_{\text{loc}}^{\left(J\right)}(\br) - E_{\text{loc}}^{(V)}(\br) \right)
                + \overline{\beta}_I^{(V)},
     \label{eq:estimatorII_def}
\end{equation}
where the control variate integrals
\begin{equation}
    \overline{\beta}_I^{(V)} = \mel**{D_I}{\Hv}{\Phi}
    \label{eq:estimatorII_cv}
\end{equation}
are analytically known, \textit{i.e.} with zero statistical fluctuations.

We have first used the pure Coulomb potential for the dressing:
\begin{equation}
     V_{1}(r_{ij}) = \frac{1}{r_{ij}}.
     \label{eq:coul_pot}
\end{equation}
This boils down to using the usual Coulomb integrals as a reference.
If no Jastrow factor is used, $\hat{H}_J = \hat{H}$ and
$\overline{{a}}_I^{(V)}(\br) = \overline{\beta}_I^{(V)}$ with zero variance.
Therefore, we expect the magnitude of the fluctuations of $\overline{{a}}_I^{(V)}$
to increase with the complexity of the Jastrow factor, even though the fluctuations
are expected to remain small.

Although the estimator of Eq. \eqref{eq:estimatorII_def} has lower fluctuations than the bare estimator
(Eq.~\eqref{eq:estimatorI_def}),
one can notice that, as long as the Jastrow factor $J(\br)$ satisfies the cusp conditions,
the local energy $E_{\text{loc}}^{\left(J\right)}(\br)$ does not diverge when $r_{12}\rightarrow 0$.
Therefore the introduction of the usual local energy $E_{\text{loc}}^{(V)}(\br)$ in $\overline{{a}}_I^{(V)}(\br)$
introduces a divergence when $r_{12}\rightarrow 0$ due to the bare Coulomb potential.
To eliminate this problem, one can simply replace the bare Coulomb potential by a repulsive non-divergent potential
and we propose here to use the long-range component of the potential commonly used in range-separated
\ac{DFT}:~\cite{toulouse_2004}
\begin{equation}
    V_{2}(r_{ij}) = \frac{\erf \left( \mu \, r_{ij} \right)}{r_{ij}}.
    \label{eq:rsdft_pot}
\end{equation}
As such a potential is used here only to reduce the fluctuations of $\overline{{a}}_I^{(V)}(\br)$,
the choice of the parameter $\mu$ is arbitrary and does not introduce any bias,
so it can be optimized to minimize the fluctuations of
$\overline{{a}}_I^{(V)}$.
As opposed to range-separated \ac{DFT} where the common choice for
$\mu$ is 1/2, here we use large values of $\mu$ (typically larger than
$10$),
such that the model potential becomes close to the Coulomb potential,
but without the divergence. As expected, we have observed that $V_{2}$
further reduces the fluctuations of the sampled quantities.

\subsubsection{Further improved estimator}
\label{subsec:more_improved}

In order to propose a further improved estimator we introduce an auxiliary Jastrow factor
$\mathcal{J}$
which depends on a set of parameters $\{ \mathbf{p} \}$,
such that the corresponding \ac{TC} Hamiltonian
\begin{equation}
    \hat{H}_{\mathcal{J}} \equiv e^{-\mathcal{J}\left( \mathbf{p} \right) }
                            \, \hat{H} \, e^{\mathcal{J}\left( \mathbf{p} \right) } ,
\end{equation}
allows a deterministic calculation for the integrals in the basis of determinants.
With such a Jastrow factor, we can define the \ac{ZV} estimator
\begin{equation}
    \overline{\overline a}_I(\br) = \frac{D_I(\br)}{\Phi(\br)} \,
    \left( E_{\text{loc}}^{\left(J\right)}(\br) - E_{\text{loc}}^{\left( \mathcal{J} \right)}(\br) \right)
                + \overline{\overline \beta}_I,
    \label{eq:estimatorIII_def}
\end{equation}
with
\begin{equation}
     E_{\text{loc}}^{\left( \mathcal{J} \right)}(\br) = \frac{ \hat{H}_{\mathcal{J}} \, \Phi(\br) }{\Phi(\br)},
\end{equation}
and
\begin{equation}
    \overline{\overline \beta}_I = \mel**{D_I}{ e^{-\mathcal{J}\left( \mathbf{p} \right) }
    \, \hat{H} \, e^{\mathcal{J}\left( \mathbf{p} \right) } }{\Phi}.
\end{equation}
While the Jastrow factor $\mathcal{J}$ is selected to mimic $J$, by optimizing the
parameters $\mathbf{p}$, the statistical fluctuations of the
estimators of interest can be significantly
reduced.~\cite{assaraf_1999}

The control variate quantities $\overline{\overline \beta}_I$ can be
prohibitively expensive to calculate, even in close form, due to the three-electrons
terms (which generate a six-index tensor) inherent to any \ac{TC} Hamiltonian. To avoid this complexity, we introduce the modified
local energy
\begin{equation}
    \epsilon_{\text{loc}}^{\left( \mathcal{J} \right)}(\br) =
    E_{\text{loc}}^{\left( \mathcal{J} \right)}(\br) - E_{3-\text{e}}^{\left( \mathcal{J} \right)}(\br)
\end{equation}
where we have dropped the three-electron contributions involved in the
computation of the local energy $E_{3-\text{e}}^{\left( \mathcal{J} \right)}$.
Therefore, the improved estimator becomes
\begin{equation}
    \tilde{a}_I(\br) = \frac{D_I(\br)}{\Phi(\br)} \,
    \left( E_{\text{loc}}^{\left(J\right)}(\br) - \epsilon_{\text{loc}}^{\left( \mathcal{J} \right)}(\br) \right)
                + \tilde{\beta}_I ,
    \label{eq:estimatorIV_def}
\end{equation}
with
\begin{equation}
    \tilde{\beta}_I  = \mel**{D_I}{ e^{-\mathcal{J}\left( \mathbf{p} \right) }
    \, \hat{H} \, e^{\mathcal{J}\left( \mathbf{p} \right) }- E_{3-\text{e}}^{\left( \mathcal{J} \right)}}{\Phi}.
\end{equation}
Note that the estimator of Eq. \eqref{eq:estimatorIV_def} is not biased since the three-electron terms
arising from the Jastrow factor $J(\br)$ are taken into account in $\tilde{a}_I(\br)$,
and the use of $ \epsilon_{\text{loc}}^{\left( \mathcal{J} \right)}(\br)$ in Eq. \eqref{eq:estimatorIV_def}
is only here to reduce the fluctuations.
Therefore, the set of parameters $\{\mathbf{p}\}$ can still be optimized to minimize
the variance of the dressing elements.
One way to do that is, for instance, through minimizing
the sum of the variance of integrals weighted with the squared \ac{CI} coefficients:
\begin{equation}
    \sum_{I=1}^{\Ndet} c_I^2 \, \Var \qty(\tilde{a}_I)
    = \Var \left( E_{\text{loc}}^{\left(J\right)} - \epsilon_{\text{loc}}^{\left( \mathcal{J} \right)} \right).
    \label{eq:wight_var}
\end{equation}

\subsubsection{Choice of $\mathcal{J}$ for further improved estimators}

Several correlation factors that allow a deterministic calculation of
integrals have been proposed in the literature of \ac{TC}
methods.~\cite{Nooijen_1998,ten_2000_I,ten_2000_II,Giner_2021}
We chose here to consider a two-electron Jastrow factor accounting for the
cusp condition and the short-range part of the Coulomb hole,
together with a one-body Jastrow factor which allows to compensate for
the effect of the two-body Jastrow on the one-body density.
Therefore, the general form of such a Jastrow factor reads
\begin{equation}
    \mathcal{J} = \sum_{i<j}^N \, u \left( r_{ij}; \mu \right) -
    \sum_{i=1}^N \sum_{A=1}^M \,  v \left( r_{iA}; \beta_A \right),
\end{equation}
where, $M$ is the number of nuclei, $r_{iA}$ is the distance between
the $i-$th electron and the $A-$th nucleus, and $\{ \beta_A \}$ are some positive parameters.

Regarding the two-electron Jastrow factor, we used the recently proposed correlation factor
tuned by a single-parameter~\cite{Giner_2021}
\begin{equation}
    u \left( r_{ij}; \mu \right) =
    \frac{r_{ij}}{2} \left[ 1 - \erf \left( \mu r_{ij} \right) \right]
    - \dfrac{\exp \left[ \left( \mu r_{ij} \right)^2 \right]}{2 \sqrt{\pi} \, \mu} ,
    \label{eq:u_Manu}
\end{equation}
which imposes the electron-electron cusp conditions and whose corresponding \ac{TC} Hamiltonian
reproduces the effective Hamiltonian of RSDFT at leading-order in $1/r_{ij}$.
The explicit form of the \ac{TC} Hamiltonian obtained with such a Jastrow factor was given in Ref. \onlinecite{Giner_2021}.

Turning now to the one-body Jastrow factor chosen in the present work, its functional form reads
\begin{equation}
    v \left( r_{iA}; \beta_A \right) = 1 - \exp \left( - \beta_A \, r_{iA}^2 \right) .
\end{equation}

Within these definitions, the \ac{TC} Hamiltonian corresponding to $\mathcal{J}$ is given by
\begin{equation}
 \begin{aligned}
 \hat{H}_{\mathcal{J}}  = \hat{H} & - \sum_{i} \vone{i} - \sum_{i<j} \big( \kmu{i}{j} + \kone{i}{j} \big)  \\
                            & - \sum_{i<j<k} \lmu{i}{j}{k},
 \end{aligned}
\end{equation}
where the analytical expressions of the additional terms with respect to $\hat{H}$ are
\begin{equation}
  \begin{aligned}
    \vone{i} = &- \sum_A \beta_A\, e^{- \beta_A \, r_{iA}^2}
     \left[  3 - 2\, \beta_A\, r_{iA}^2 + 2 \left(  \brs{i} - \mathbf{R}_A \right) \cdot \grad_i \right]  \\
		       &+ 2 \left[ \sum_A \beta_A\, e^{- \beta_A \, r_{iA}^2} \left(  \brs{i} - \mathbf{R}_A \right) \right]^2,
	\end{aligned}
\end{equation}
\begin{equation}
 \label{eq:k_final}
  \begin{aligned}
    & \kmu{i}{j}=  \frac{1 - \text{erf}(\mu \, r_{ij})}{r_{ij}} -
    \frac{\mu}{\sqrt{\pi}} e^{-\big(\mu \, r_{ij} \big)^2} \\
    &+ \frac{\bigg(1 -     \text{erf}(\mu \, r_{ij})
      \bigg)^2}{4} - \bigg( \text{erf}(\mu \, r_{ij}) - 1\bigg) \deriv{}{r_{ij}}{},
  \end{aligned}
\end{equation}
\begin{equation}
  \begin{aligned}
    &\kone{i}{j} = - \left( \frac{1 - \erf(\mu \, r_{ij})}{r_{ij}} \right) \sum_A \beta_A  \\
    & \times \left( \brs{i} - \brs{j}\right) \cdot
    \left[  \left(  \brs{i} - \mathbf{R}_A \right)  e^{- \beta_A \, r_{iA}^2}
      - \left(  \brs{j} - \mathbf{R}_A \right) e^{- \beta_A \, r_{jA}^2} \right],
  \end{aligned}
\end{equation}
and
\begin{equation}
 \label{eq:l_final}
 \begin{aligned}
   \lmu{i}{j}{k} = & \frac{1 - \text{erf}(\mu \, r_{ij})}{2 r_{ij}}
   \brs{ij} \cdot \frac{1 - \text{erf}(\mu \, r_{ik})}{2 r_{ik}} \brs{ik} \\
   + & \frac{1 - \text{erf}(\mu \, r_{ji})}{2 r_{ij}} \brs{ji}
   \cdot \frac{1 - \text{erf}(\mu \, r_{jk})}{2 r_{jk}} \brs{jk} \\
   + & \frac{1 - \text{erf}(\mu \, r_{ki})}{2 r_{ik}} \brs{ki}
   \cdot \frac{1 - \text{erf}(\mu \, r_{kj})}{2 r_{kj}} \brs{kj}.
 \end{aligned}
\end{equation}
The computation of analytical or mixed analytical integrals on a Gaussian basis for $\kmu{i}{j}$ and $\lmu{i}{j}{k}$
have been given in Ref. \onlinecite{Giner_2021}, and similar schemes have been used for the integrals involving $\vone{i}$ and $\kone{i}{j}$.

\section{Numerical results}\label{sec:opt_numeric}
In the present section we investigate the efficiency of the \ac{TC}
approach by performing a series of test calculations on both atomic and molecular systems.
The wave function calculations were performed with the \textsc{Quantum
  Package}~\cite{garniron_2019} program,
and the \ac{QMC} calculations were made with \textsc{QMC=Chem}.~\cite{scemama_2013,scemama_2013_2}
A plug-in was developed in \textsc{Quantum Package} to read the elements
sampled with \textsc{QMC=Chem} and add the deterministic control variate quantity for the dressing of the Hamiltonian to be diagonalized.
Throughout this work, the initial wave function used in the iterative dressing is the ground state eigenvector of the usual Hamiltonian
and the basis of Slater determinants is kept fixed.

In Sec.~\ref{sec:Be} we report a detailed study on the Be atom in order to compare the present optimization scheme
with the usual linear method and stochastic reconfiguration.
Then, in Sec.~\ref{sec:mol}  we perform more realistic calculations with wave functions containing
several hundreds of thousands of Slater determinants on atomic and molecular systems.

\subsection{The Beryllium atom}
\label{sec:Be}
In the present section we use a small system, the Be atom, in order to investigate several numerical aspects of the present work:
i) the convergence of the iterative scheme described in Sec.~\ref{sec:iterative} used to obtain the right-eigenvector of a given non-Hermitian matrix,
and ii) the effect of the incompleteness of the basis set on the discrepancy between the true variational minimum of $\hvmcb$ and $\htcb$ as discussed in Sec.~\ref{sec:tc_for_vmc}.
The numerical study of the reduction of the statistical fluctuations of the present scheme through
improved estimators will be presented in Sec.~\ref{sec:mol}.
Throughout Sec.~\ref{sec:Be}, we project both the \ac{TC} and \ac{VMC} Hamiltonians on the $N$-electron basis set $\basis$
made of the Hartree-Fock determinant in the cc-pCVDZ atomic basis set and all singly and doubly excited determinants,
which results in a set of about 350 determinants.
Also, we use the following Jastrow factor $J(\br)$ to define both the \ac{TC} and \ac{VMC} Hamiltonians
\begin{equation}
 \label{eq:j_mu_as_j}
    J(\br) = \sum_{i<j} \, u \left( r_{ij}; \mu \right),
\end{equation}
with $\mu=1.0$.
In the case of the \ac{TC} Hamiltonian, thanks to the simple
analytical form of $J(\br)$, the two- and three-body integrals
involved in the \ac{TC} Hamiltonian can be computed exactly, and we
therefore avoid any problems related to the stochastic sampling
inherent to \ac{VMC}.

\subsubsection{Iterative scheme to obtain right-eigenvector of the TC Hamiltonian}
To analyze the iterative scheme leading to the lowest right-eigenvector of the \ac{TC} Hamiltonian,
we built explicitly the matrix of the \ac{TC} Hamiltonian within
$\basis$ and obtained as a reference the exact \ac{TC} ground state
eigenvalue and eigenvectors within $\basis$ by using a non-Hermitian
eigensolver present in the LAPACK~\cite{lapack} library.

\begin{figure*}[htp]
  \subfloat[\label{fig:dress_Be_Etc}]{
    \includegraphics[width=0.45\linewidth]{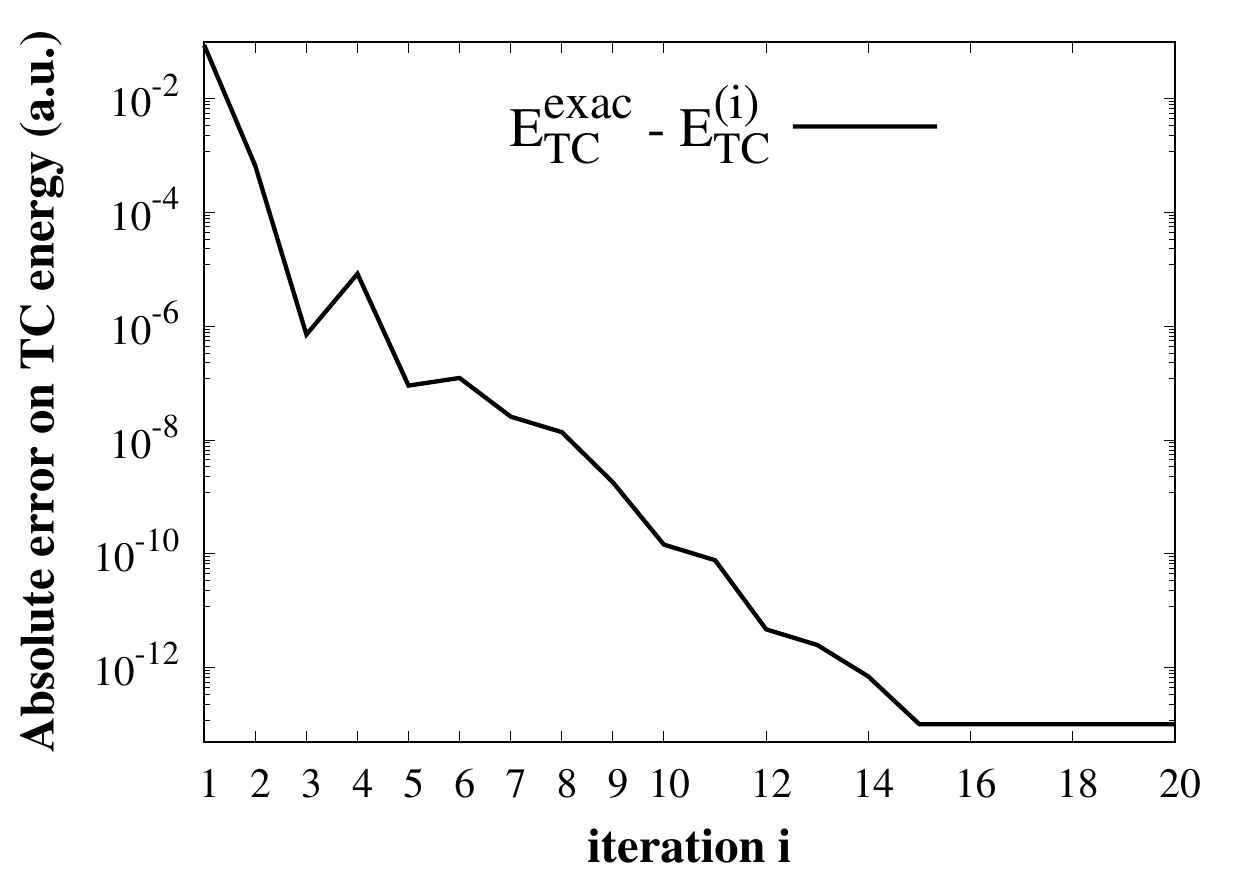}
  } \quad
  \subfloat[\label{fig:dress_Be_Evmc}]{
    \includegraphics[width=0.50\linewidth]{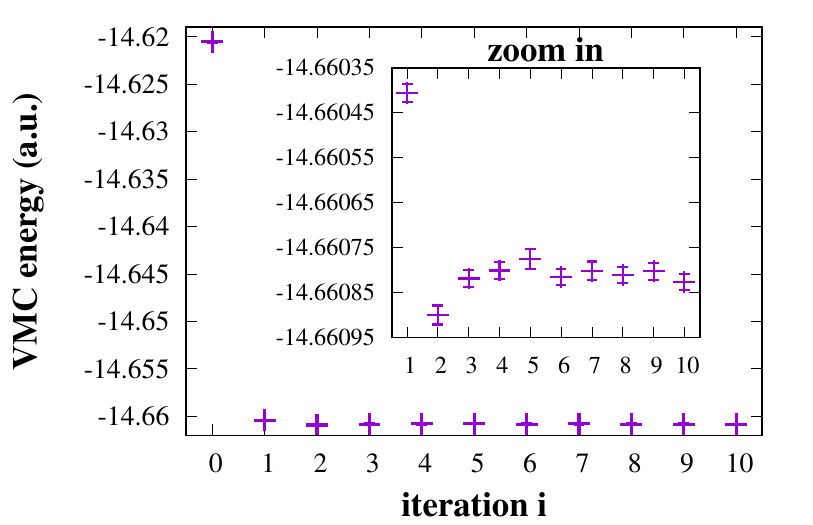}
  }
  \caption{Convergence of the \ac{TC} and \ac{VMC} energies along
    the iterative dressing.
    (a) Difference between the \ac{TC} energy obtained by
    a direct diagonalization and that obtained by the iterative dressing technique.
    (b) \ac{VMC} energy of the \ac{CI}-J wavefunction before optimizing ($i=0$) and
    during the iterative dressing.
  }
  \label{fig:dress_Be}
\end{figure*}

We report in Fig.~\ref{fig:dress_Be} the convergence of the absolute error between the exact ground state eigenvalue and that obtained with the iterative scheme at a given iteration.
From Fig.~\ref{fig:dress_Be} one can notice that the iterative scheme converges, although not in a monotonic way for the first five iterations, towards the exact energy.
A detailed analysis of the data shows that an error of about $\sim 0.6$~mH is reached in only two iterations.
This experiment validates that without statistical fluctuations, this method returns the right eigenvector of the
\ac{TC} Hamiltonian within chemical accuracy in about $2-3$ iterations.
This observation appears to be true for more complex systems as it will be shown in the next sections.

One can now focus on the efficiency of the \ac{TC} iterative eigensolver with respect to the optimization
in the sense of the \ac{VMC} energy (\textit{i.e.} in the presence of the Jastrow factor of Eq.~\eqref{eq:j_mu_as_j}).
We reported in Fig.~\ref{fig:dress_Be} the convergence of the \ac{VMC} energy of the wave function at a given iteration
of the iterative scheme.
From Fig.~\ref{fig:dress_Be} it appears that the first two iterations lower the \ac{VMC} energy by about $\sim 40$~mH,
the successive iterations having only a minor impact on the \ac{VMC} energy are within two statistical deviations.

\subsubsection{Effect of an incomplete basis set and comparison with other optimization methods}
Having validated the approach of the iterative dressing technique to obtain the eigenvector of the \ac{TC} Hamiltonian in a given Slater determinant
basis set $\basis$, we propose here to investigate the effect of the incompleteness of such a basis set.
Indeed, as shown in the appendix (see Sec.~\ref{sec:appendix}), the ground state eigenvector of
$\hvmcb$ (\textit{i.e.} the \ac{VMC} Hamiltonian projected in $\basis$) does not necessarily
coincide with the ground state right-eigenvector of $\htcb$ (\textit{i.e.} the \ac{TC} Hamiltonian projected in $\basis$) as long as $\basis$ is
incomplete. Therefore the variational energy obtained with the latter is necessarily an upper bound to the ground state eigenvalue of $\hvmcb$,
and we propose here to quantify this on a simple system such as the Be atom.

\begin{figure*}[htp]
  \subfloat[\label{fig:Be_GEP}]{
    \includegraphics[width=0.48\linewidth]{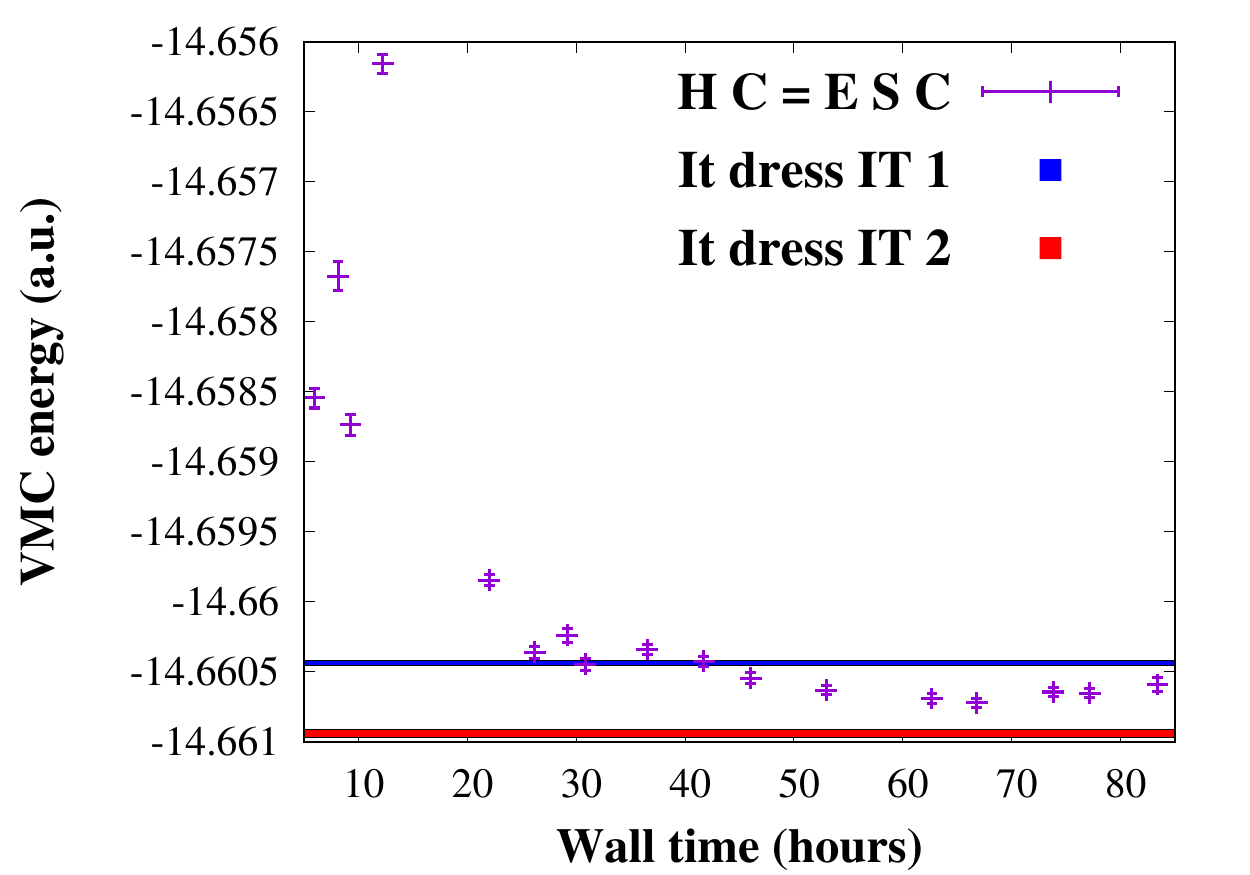}
  } \quad
  \subfloat[\label{fig:Be_SR}]{
    \includegraphics[width=0.48\linewidth]{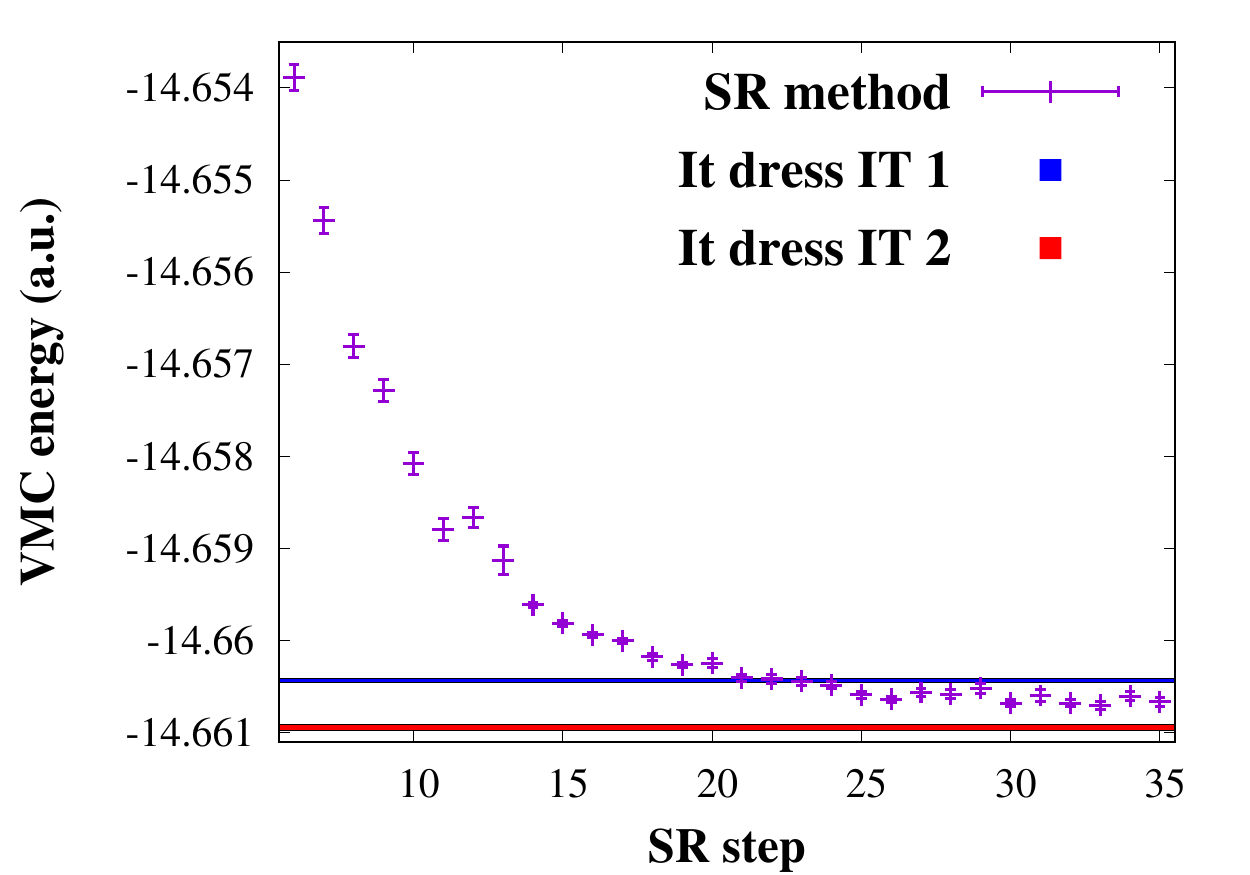}
  }
  \caption{Convergence of the variational energies of the ground state eigenvector of
    (a) the \ac{GEP} method and (b) the SR method. On each figure we report the variational
    energies obtained with the two first iterations of the dressing scheme.
  }
  \label{fig:Be_vmc}
\end{figure*}

We sampled the matrix elements of $\hvmcb$ together with the corresponding overlap matrix and solved
the \ac{GEP} of the form of Eq.~\eqref{eq:general_h} to obtain the ground state
eigenvector. The matrix elements are impacted by statistical
fluctuations, so we have made simulations with increasingly large
numbers of samples. To show the impact of these fluctuations on the
energy of the obtained eigenvector,
we report in Fig~\ref{fig:Be_vmc} the convergence of the variational energy of the
eigenvector as a function of the wall-clock time used to sample the
matrices with 72 CPU cores.
We also report on the same figure the convergence of the \ac{SR} approach applied to the same problem with a
step size $\tau = 0.01$ (Eq.\eqref{eq:SR_lineq}).
We compare these variational energies with those obtained with the two first iterations of the iterative dressing scheme.
In order to make the comparison more fair, instead of using the exact analytical integrals involved in $\htcb$,
we sampled the dressing elements in \ac{VMC} using the
estimator $\overline{{a}}_I^{(V_1)}$ of Eq.~\eqref{eq:estimatorII_def}
with the Coulomb potential of Eq.~\eqref{eq:coul_pot}.
At each iteration, the dressing vector was sampled using a run of 50 minutes on 72 cores.
From Fig.~\ref{fig:Be_vmc} we can observe that the variational
energies obtained with the two schemes converge essentially to the
same energy as the one
obtained with the \ac{TC} optimization scheme, the latter being the lowest obtained within the CPU time spent.
One can also notice the slow convergence and the erratic behaviour of the \ac{GEP} approach,
which is caused by the large fluctuations of the sampled matrix elements~\cite{nightingale_2001}.
Regarding the SR approach, one can also see the slow rate of convergence.
These calculations show that, even for a basis of Slater determinants far from being complete such as the \ac{CISD} in a cc-pCVDZ basis set, the error
with respect to the true minimum is negligible, and in that precise case the new scheme provides a lower variational energy than
the two other schemes based on usual \ac{VMC} approaches.
This demonstrates that the error induced by a finite sampling are more important than the finite basis approximations.

\subsection{Large wave function optimizations on atomic and molecular systems}
\label{sec:mol}
The optimization of \ac{CI} coefficients in all-electron calculations
is considered a difficult task because of the wide fluctuations of
the sampled matrix elements. The large magnitude of the contributions
to the local energy of the electrons close to the nucleus are
responsible for these large fluctuations.
As illustrated in Fig.~\ref{fig:Be_vmc}, this strongly impacts the
rate of convergence of the optimization algorithms.
Here, we illustrate the efficiency of our
improved estimators on all-electron calculations of \ce{C2}, \ce{N2},
\ce{O2}, and \ce{F2} with expansions made of a few hundred thousand parameters.
It is worth mentioning that point-group symmetry was not exploited in the optimization.

Throughout this section, the following Jastrow factor was used:
\begin{equation}
    J(\br) = \sum_{i<j}^N \frac{ r_{ij}}{2(1+b \, r_{ij}) }
              - \sum_{i=1}^N \sum_{A=1}^M  \left( \frac{\alpha_A \, r_{iA}}{1+\alpha_A \, r_{iA} } \right)^2.
    \label{eq:J_simple}
\end{equation}
We performed all electrons calculations using Dunning's cc-pVTZ basis set.~\cite{dunning_1989},
and in order to investigate the behaviour of the present schemes within pseudo-potential calculations, we report calculations in the case of the 
\ce{N2H4} molecule using the Burkatzki-Filippi-Dolg \acp{ECP}~\cite{burkatzki_2007}
in the compatible double zeta basis set. The energy was computed in the determinant localization approximation.\cite{zen_2019}

For all diatomics, the bond lengths were taken to be the experimental
ones given by Huber and Herzberg.\cite{Huber_1979} The geometry of \ce{N2H4}
is the experimental geometry.\cite{} 
The Jastrow factors were optimized at the \ac{HF} level with a single determinant, and then
frozen-core CIPSI calculations were made in the full valence \ac{CI}
space using Hartree-Fock orbitals in the cc-pVTZ basis set to generate
the initial \ac{CI} expansions.

\subsubsection{Reduction of statistical fluctuations through control variates}
\label{sec:red_fluct}

\begin{figure}[tp]
    \includegraphics[width=\linewidth]{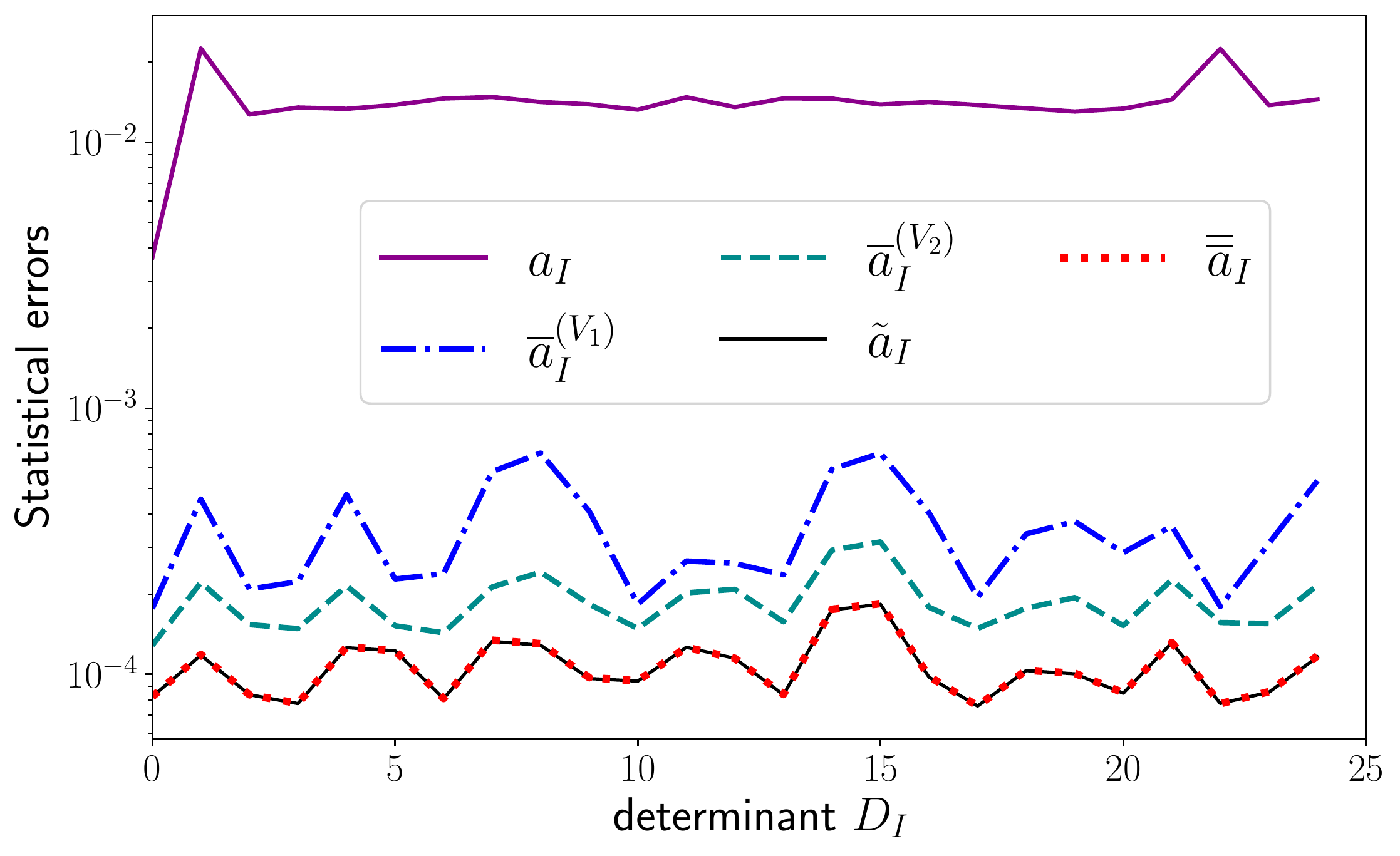}
    \caption{Statistical errors of the estimators:
      $a_I$~(Eq.~\eqref{eq:estimatorI_def}),
      $\overline{{a}}_I^{(V_1)}$ and $\overline{{a}}_I^{(V_{2})}$ (Eq.~\eqref{eq:estimatorII_def}),
      $\overline{\overline a}_I$ (Eq.~\eqref{eq:estimatorIII_def}),
      and $\tilde{a}_I$ (Eq.~\eqref{eq:estimatorIV_def}),
      obtained in a common \ac{VMC} sampling.
    }
    \label{fig:stat_err}
\end{figure}

We begin this study by investigating the reduction of statistical fluctuations of the sampled quantities using
the various control variates presented in Sec.~\ref{subsec:dress_elements}.
We report in Fig.~\ref{fig:stat_err} a comparison between the statistical errors obtained in a single \ac{VMC} run
for the 25 most important determinants of the \ce{N2} molecule using different estimators:
i)  the standard estimator $a_I$ given in Eq.~\eqref{eq:estimatorI_def};
ii) the improved estimators $\overline{{a}}_I^{(V_1)}$ and $\overline{{a}}_I^{(V_{2})}$
of Eq.~\eqref{eq:estimatorII_def}, corresponding to the model
potentials of Eqs.~\eqref{eq:coul_pot} and \eqref{eq:rsdft_pot} with $\mu=10$, respectively;
iii) the \ac{TC}-correlated estimators
$\overline{\overline a}_I$ and $\tilde{a}_I$ (Eqs.~\eqref{eq:estimatorIII_def}
and~\eqref{eq:estimatorIV_def}) which involve the Jastrow factor $\mathcal{J}$
with the parameters given in table~\ref{tab:dress_res}.
For \ce{N2}, as well as for other systems considered in the following,
the parameters were optimized to reduce the fluctuations of the weighted
variance given in Eq~\eqref{eq:wight_var}.

As apparent from Fig.~\ref{fig:stat_err}, the fluctuations of the improved estimators are
much smaller than those of the standard estimator. The singularity-free potential
(estimator $\overline{{a}}_I^{(V_2)}$) yields smaller statistical errors than
the Coulomb potential (estimator $\overline{{a}}_I^{(V_1)}$). The noise is further reduced
using the \ac{TC}-based estimators $\overline{\overline a}_I$ and $\tilde{a}_I$,
and we can observe that neglecting the three-electron terms in the
local energy does not increase the fluctuations.
The latter is extremely important from a computational perspective as the \ac{TC} calculations
without any of the three-electron terms has essentially the same
computational scaling as a standard calculation
using the regular Hamiltonian (\textit{i.e.} without any Jastrow factor).

To get an overview of the reduction of the statistical error on the complete set
of coefficients, we compared for each estimator $a$ the ratio of the
sum of the standard deviations with the one obtained with the bare
estimator $\alpha$ using the ratio
\begin{equation}
  r(a) = \frac{\sum_{I=1}^{\Ndet} \sigma(a_I)}{\sum_{I=1}^{\Ndet} \sigma(\alpha_I)}.
\end{equation}
We obtained
$r(\overline{{a}}_I^{(V_1)}) = 0.027$,
$r(\overline{{a}}_I^{(V_2)}) = 0.022$,
$r(\overline{\overline a}_I) = 0.009$,
$r(\tilde{a}_I) = 0.009$, showing that the reduction of statistical
error is observed for all determinants and not only the most important
ones.

\subsubsection{Optimization of large wave functions for molecular systems}
\label{sec:opt_large}


\begin{table*}[htp]
  \caption{Parameters and energies
    (a.u.) associated with the different wave functions. The convergence
    of the energy with the iterations of optimization is shown in the
    bottom part of the table.}
  \begin{ruledtabular}
  \begin{tabular}{lddddd}
                          &  \mt{\ce{C2}} &    \mt{\ce{N2}}  &   \mt{\ce{O2}}  & \mt{\ce{F2}}  & \mt{\ce{N2H4}}                     \\
    \hline
 Basis set                & \mt{cc-pVTZ}  & \mt{cc-pVTZ}     &   \mt{cc-pVTZ}  & \mt{cc-pVTZ}  & \mt{BFD/cc-pVDZ}                   \\
 $N_\text{det}$           & \mt{287~998}  & \mt{298~250}     &   \mt{123~711}  & \mt{191~538}  & \mt{1~000~000}                     \\
 $r_0$ (\AA)              &    1.2425     &     1.0977       & 1.2075          &   1.4119      & -                                 \\
 $b$                      & 1.382         & 1.829            & 1.923           &   2.206       & 1.000                              \\
 $\alpha$                 & 0.882         & 0.691            & 0.882           &   0.691       & \mt{N:0.690, H:0.170}              \\
 $\mu$                    & 0.738         & 2.161            & 2.263           &   1.619       &  $-$                               \\
 $\beta$                  & 0.635         & 4.288\times10^{-3} & 6.209\times 10^{-3} & 0.410   &  $-$                               \\
    HF energy             &  -75.40144    &  -108.98347      &  -149.65257     & -198.75204    &   -21.73877                         \\
    HF/J  energy          &  -75.6587(1)  &  -109.2668(1)    &  -150.0060(1)   & -199.0884(1)  &   -22.1008(1)                       \\
    Selected CI energy    &  -75.77204    &  -109.35465      &  -150.09752     & -199.26114    &   -22.11353                         \\
    \hline
 Energies along the optimization: \\
    \hline
    CI/J IT 0             & -75.8808(1)  &  -109.4785(1)    &  -150.2451(2)   & -199.3766(3)  & -22.1969(5)                        \\
    CI/J IT 1             & -75.8896(1)  &  -109.4986(1)    &  -150.2670(1)   & -199.4297(2)  & -22.2297(3)                        \\
    CI/J IT 2             & -75.8902(1)  &  -109.4986(1)    &  -150.2671(1)   & -199.4300(2)  & -22.2297(3)                        \\
    CI/J IT 3             & -75.8903(1)  &  -109.4989(1)    &  -150.2673(1)   & -199.4306(2)  & -22.2302(3)                        \\
  \end{tabular}
  \end{ruledtabular}
  \label{tab:dress_res}
\end{table*}

Having established the estimator with the lowest statistical fluctuations in Sec.~\ref{sec:red_fluct},
we optimize the \ac{CI} parameters of the CIPSI-Jastrow wave functions using the estimator
$\tilde{a}_I(\br)$ given in Eq.~\eqref{eq:estimatorIV_def}.
We begin our study with all electron calculations on the \ce{C2}, \ce{N2}, \ce{O2} and \ce{F2} molecules.
The parameters of $\mathcal{J}$ for each molecule and the corresponding
\ac{VMC} energies are reported in table~\ref{tab:dress_res}.

For all systems, we observe a sub-milliHartree convergence after two
optimization iterations, although a single iteration could be sufficient.
The gains in \ac{VMC} energy are 10~mH for \ce{C2}, 20~mH for
\ce{N2}, 22~mH for \ce{O2} and 53~mH for \ce{F2}.
It is remarkable that the optimization is still stable for the \ce{F2} molecule which
has the largest total energy. These results confirm that this
optimization method can be used as a black box method in routine calculations.

We conclude our study by a calculation on the
N$_2$H$_4$ molecule using a BFD \ac{ECP} with
the corresponding double zeta basis set.
For this system, $1~000~000$ CI coefficients are optimized.
The results of the optimization are presented in table~\ref{tab:dress_res} and show that two iterations are
required to converge to the minimum energy which is about 
$\sim 32$~\si{\milli\hartree} below the initial energy. This result confirms that the
present scheme allows the optimization in an \ac{ECP} framework of very large
CI expansions even with the simplest estimator (Eqs.~\eqref{eq:estimatorII_def} and \eqref{eq:coul_pot}).

\section{Summary}
\label{sec:conclu}
We have presented an iterative method to optimize large \ac{CI} expansions in
the presence of a general correlation factor. The main idea is to consider a
similarity-transformation of the Hamiltonian by the Jastrow factor
which results in an effective Hamiltonian, the so-called \ac{TC} Hamiltonian,
having the same right-eigenvectors than the usual ground state \ac{VMC} eigenvector in the limit of a complete basis.
The effect of the Jastrow factor in the \ac{TC} approach can be written as an additive dressing of the standard Hamiltonian.
The \ac{QMC} simulations are then used only to sample the quantities required for such a dressing, and
to compute the variational energy associated with the wave function.
The main advantages of the present approach are i) a large part of the
quantities required to optimize the wave function are analytical
(\textit{i.e.} with zero fluctuations), ii) the number of quantities
to sample is equal to the number of determinants, and iii)
the fluctuations of the sampled quantities are small.

After having illustrated in Sec.~\ref{sec:Be} the robustness of the present approach on calculations on the Be atom
even far from the complete basis set limit (\textit{i.e.} in realistic cases),
we have shown in Sec.~\ref{sec:mol} its efficiency on wave functions made of a few hundred thousand Slater determinants.
We demonstrated in Sec.~\ref{sec:red_fluct} how one can significantly
lower the statistical fluctuations of the computed quantities thanks
to control variates, and in Sec.~\ref{sec:opt_large} how it performs
on molecular systems with wave functions with 
$10^5 - 10^6$ determinants. The efficiency of this approach comes from the
mixing of the deterministic transcorrelated method,
with the \ac{QMC} method: a large fraction of the needed matrix elements
can be computed in the standard framework of wave function methods, and only a
small number of contributions come from the \ac{QMC} simulations. Therefore,
the quantities of interest have by nature a very low variance. It is possible
to put more computational weight on the deterministic part of the calculation
to reduce even more the statistical fluctuations, and the user has the
flexibility to find the best compromise between the computational cost of the
control variates and the \ac{QMC} simulations.

We have also shown that this method has several advantages regarding computational
considerations. First, the memory required in the \ac{QMC} code is minimal:
only $\Ndet$ quantities need to be stored since the overlap matrix is not
needed, and these quantities are computed using building blocks that are already
needed for the computation of the energy. Hence, the extra computational cost is
also minimal. Finally, it is important to mention that the fast convergence of the
method (less than three iterations) is extremely important when considering
massively parallel simulations. The only necessary blocking communications
take place at the beginning and at the end of an iteration, so it is preferable to
have an optimization algorithm with long computing phases and very few
iterations than the opposite.

In this paper, we have limited our examples to the optimization of \ac{CI}
expansions, but it is worth mentioning that this method can of course be
applied in a super-\ac{CI} framework for the optimization of the coefficients of the
molecular orbitals.
We plan to elaborate more on these aspects in a subsequent work.

\begin{acknowledgments}
This work was performed using HPC resources from GENCI-TGCC
(2021-gen1738) and from CALMIP (Toulouse) under allocation
P22001, and was also supported by the European Centre of
Excellence in Exascale Computing TREX --- Targeting Real Chemical
Accuracy at the Exascale. This project has received funding from the
European Union's Horizon 2020 --- Research and Innovation program ---
under grant agreement no.~952165.
\end{acknowledgments}




\section{Appendix}
\label{sec:appendix}
\subsection{VMC, Transcorrelation and projection in an incomplete basis set}
The aim of the present section is to make a proper link between the \ac{VMC} optimization and the \ac{TC} approach when they are both projected in the same basis set.
\subsubsection{Definition of the projector on a basis set}
Let $\basis$ be a $N$-electron basis set, which can be for instance a set of $N$-electron Slater determinant
\begin{equation}
 \basis = \{ \ket{\rm I},{\rm I}=1,N_\basis \},
\end{equation}
which we will assume orthonormal for the sake of simplicity.
Such a basis span a vector space $\vbasis$ which is made of any functions $\ket{\Psi_\basis}$ being a linear combinations
of the Slater determinants in $\basis$
\begin{equation}
 \vbasis = \bigg\{ \ket{\Psi_\basis} = \sum_{\rm I \in \basis} c_{\rm I} \ket{\rm I}, \,\, c_{\rm I} \in \mathbb{C}\bigg\}.
\end{equation}
From $\basis$ one can build the projector $\pbasis$
\begin{equation}
 \pbasis = \sum_{\rm I} \ket{\rm I} \bra{\rm I},
\end{equation}
which verifies the projector property
\begin{equation}
 \label{eq:def_proj}
 \begin{aligned}
& \pbasis \pbasis = \pbasis.
 \end{aligned}
\end{equation}
Such a projector coincide with the identity for all wave functions in $\vbasis$
\begin{equation}
 \pbasis \ket{\Psi_\basis} = \ket{\Psi_\basis}.
\end{equation}
Therefore, in virtue of Eq.\eqref{eq:def_proj} one obtains that
\begin{equation}
 \pbasis \pbasis \ket{\Psi_\basis} = \ket{\Psi_{\basis}},
\end{equation}
which implies that
\begin{equation}
 \big(\pbasis\big)^{-1} = \pbasis.
\end{equation}

One can define the \textit{complementary basis set} $\obasis$ which is the set of Slater determinants such that it completes the basis set
\begin{equation}
 \obasis = \{ \ket{\rm L} \notin \basis\},
\end{equation}
the corresponding vector space $\vobasis$
\begin{equation}
 \vobasis = \bigg\{ \ket{\Psi_\obasis} = \sum_{{\rm L} \in \obasis} c_{\rm L} \ket{\rm L} \bigg\}
\end{equation}
and the corresponding projector $\pobasis$ such that
\begin{equation}
 \label{eq:def_identity}
 \pbasis + \pobasis = \ident ,
\end{equation}
where $\ident$ is the identity operator defined on the complete basis made of the reunion of $\vbasis$ and $\vobasis$
\begin{equation}
 \ident \ket{\Psi} = \ket{\Psi} \quad \forall \ket{\Psi} \in \vbasis \cup \vobasis.
\end{equation}
An important property is that the projector $\pbasis$ and the complementary projector $\pobasis$ are orthogonal
\begin{equation}
 \label{eq:pb_pob}
 \pbasis \pobasis= \pobasis \pbasis = 0.
\end{equation}

\subsubsection{Link between the VMC and transcorrelated Hamiltonian within the same basis set }
The purpose of the present section is to establish the formal link between the \ac{VMC} Hamiltonian and the \ac{TC} one projected in a basis set.
Let $J$ be a Jastrow factor, one can then define the \ac{VMC} Hamiltonian as
\begin{equation}
 \hvmc = e^{J} H e^{J},
\end{equation}
the corresponding \ac{TC} Hamiltonian as
\begin{equation}
 \htc= e^{-J} H e^{J},
\end{equation}
which are of course related by
\begin{equation}
 \hvmc = e^{2J} \htc.
\end{equation}
The corresponding operators projected onto a basis set $\basis$ are defined as
\begin{equation}
 \hvmcb = \pbasis \hvmc \pbasis,
\end{equation}
\begin{equation}
 \htcb = \pbasis \htc \pbasis,
\end{equation}
and we would like to express $\hvmcb$ in terms of $\htcb$.
To do so we write
\begin{equation}
 \begin{aligned}
 \hvmcb &= \pbasis e^{2J} \htc \pbasis \\
        &= \pbasis e^{2J} \ident \ident \htc \pbasis \\
        &= \pbasis e^{2J} (\pbasis + \pobasis) (\pbasis + \pobasis) \htc \pbasis \\
 \end{aligned}
\end{equation}
which, in virtue of Eq. \eqref{eq:pb_pob}, reads
\begin{equation}
 \begin{aligned}
 \hvmcb &= \pbasis e^{2J} \pbasis \htcb  + \pbasis e^{2J} \pobasis \htc \pbasis.
 \end{aligned}
\end{equation}
If we define $\big(e^{2J}\big)_{\basis} = \pbasis e^{2J} \pbasis$ and
\begin{equation}
 \Delta^{\text{TC-VMC}}_{\basis} = \pbasis e^{2J} \pobasis \htc \pbasis
\end{equation}
we can then write
\begin{equation}
 \begin{aligned}
 \hvmcb &= \big(e^{2J}\big)_{\basis}  \htcb  + \Delta^{\text{TC-VMC}}_{\basis}.
 \end{aligned}
\end{equation}
Of course, in the limit where $\basis$ is complete, one has that $\pobasis = 0$ and then $\Delta^{\text{TC-VMC}}_{\basis} = 0$.
\subsubsection{Link between the eigenvectors of the TC and VMC}
When working on a complete basis set, we know that the eigenvectors of $\hvmc$ coincides with the right-eigenvectors of $\htc$
\begin{equation}
 \begin{aligned}
 & e^{J} H e^{J} \ket{\Psi_i} = E_i e^{2J}\ket{\Psi_i} \\
\Leftrightarrow & e^{-2J} e^{J} H e^{J} \ket{\Psi_i} = E_i e^{-2J} e^{2J}\ket{\Psi_i} \\
\Leftrightarrow & e^{-J} H e^{J} \ket{\Psi^i} = E^i \ket{\Psi^i}.
 \end{aligned}
\end{equation}
We want now to find the same kind of relationship when the operators are projected in a basis set $\basis$.
We start from the eigenvalue equation for $\hvmcb$
\begin{equation}
 \begin{aligned}
 & \hvmcb \ket{\Psi^i_\basis} = E^i_\basis \big(e^{2J}\big)_{\basis} \ket{\Psi^i_\basis},
 \end{aligned}
\end{equation}
and inserting now the expression of $\hvmcb$ in terms of $\htcb$ one obtains
\begin{equation}
 \begin{aligned}
 & \bigg(\big(e^{2J}\big)_{\basis}  \htcb  + \Delta^{\text{TC-VMC}}_{\basis} \bigg) \ket{\Psi^i_\basis} = E^i_\basis \big(e^{2J}\big)_{\basis} \ket{\Psi^i_\basis} \\
 \end{aligned}
\end{equation}
and multiplying by the inverse of $\big(e^{2J}\big)_{\basis}$ (which is $\big(e^{-2J}\big)_{\basis}$) from the left it comes
\begin{equation}
 \begin{aligned}
 \htcb  \ket{\Psi^i_\basis} + \big(e^{-2J}\big)_{\basis}\Delta^{\text{TC-VMC}}_{\basis} \ket{\Psi^i_\basis} =  E^i_\basis \ket{\Psi^i_\basis}.
 \end{aligned}
\end{equation}
One can explicit the term $\big(e^{-2J}\big)_{\basis}\Delta^{\text{TC-VMC}}_{\basis}$ which gives
\begin{equation}
 \big(e^{-2J}\big)_{\basis}\Delta^{\text{TC-VMC}}_{\basis} = \pbasis
 e^{-J} \bigg( e^{-J}\pbasis e^{J} - \ident\bigg) H e^{J}\pbasis.
\end{equation}
Defining
\begin{equation}
 \delta_{\basis} = -e^{-J} \pobasis e^{J} ,
\end{equation}
one obtains
\begin{equation}
 \big(e^{-2J}\big)_{\basis}\Delta^{\text{TC-VMC}}_{\basis} = \pbasis e^{-J}\delta_\basis H e^{J}\pbasis.
\end{equation}

Therefore, the fact that the eigenvectors of $\hvmcb$ and $\htcb$ do not coincide
comes from the fact that $\delta_\basis \ne 0$ in an incomplete basis set.

\bibliography{ci_opt}

\end{document}